\documentclass[twocolumn, apj]{emulateapj}
\usepackage{natbib, amsmath}

\newcommand{\degree}{\ensuremath{^\circ}}
\def\la{\langle}
\def\ra{\rangle}
\def\ch{{\em Chandra} }

\def\edot{\dot{e}_{\textrm{turb}}}
\def\tcool{t_{\textrm{cool}}}

\def\power{\textrm{erg}\;\textrm{cm}^{-3}\,\textrm{s}^{-1}}
\def\kms{\textrm{km}\;\textrm{s}^{-1}}
\def\kevcm{\textrm{keV}\,\textrm{cm}^2}

\newcommand       \be           {\begin{equation}}
\newcommand       \ee           {\end{equation}}

\begin{document}
\title{Turbulence in Galaxy Cluster Cores:  a Key to Cluster Bimodality?}
\author{Ian J. Parrish\altaffilmark{1}, Eliot Quataert, and Prateek Sharma\altaffilmark{1}}
\shorttitle{Turbulence and the Cooling Flow Problem}
\shortauthors{Parrish, Quataert, \& Sharma}
\affiliation{Astronomy Department \& Theoretical Astrophysics Center, 601 Campbell Hall, University of California Berkeley, CA 94720, USA; iparrish@astro.berkeley.edu}
\altaffiltext{1}{Chandra/Einstein Fellow}
\begin{abstract}
  We study the effects of externally imposed turbulence on the thermal
  properties of galaxy cluster cores, using three-dimensional
  numerical simulations including magnetic fields, anisotropic thermal
  conduction, and radiative cooling. The imposed ``stirring'' crudely
  approximates the effects of galactic wakes, waves generated by
  galaxies moving through the intracluster medium (ICM), and/or
  turbulence produced by a central active galactic nucleus.  The
  simulated clusters exhibit a strong bimodality.  Modest levels of
  turbulence, $\sim 100 \, \kms \sim 10\%$ of the sound speed,
  suppress the heat-flux-driven buoyancy instability (HBI), resulting
  in an isotropically tangled magnetic field and a quasi-stable, high
  entropy, thermal equilibrium with no cooling catastrophe. Thermal
  conduction dominates the heating of the cluster core, but turbulent
  mixing is critical because it suppresses the HBI and (to a lesser
  extent) the thermal instability.  Lower levels of turbulent mixing
  ($\lesssim 100 \, \kms$) are insufficient to suppress the HBI,
  rapidly leading to a thermal runaway and a cool-core cluster.
  Remarkably, then, small fluctuations in the level of turbulence in
  galaxy cluster cores can initiate transitions between cool-core (low
  entropy) and non cool-core (high entropy) states.
\end{abstract}
\section{Introduction}\label{sec:intro}

The cooling time in the intracluster medium (ICM) of galaxy clusters
is often $\lesssim 0.1$--1 Gyr near the center of the cluster
\citep{sarazin86}. X-ray spectroscopy shows, however, that the
majority of the plasma is not in fact cooling to temperatures well
below the mean cluster temperature (e.g., \citealt{pf06}).
Understanding the processes responsible for heating and stabilizing
cluster plasmas is a central problem in galaxy formation.  Some of the
most promising energy sources include a central active galactic
nucleus (AGN) (e.g., \citealt{bt95}), thermal conduction from large
radii (e.g., \citealt{nm01}), or dynamical friction and/or turbulence
generated by the motion of substructure through the cluster (e.g.,
\citealt{kim05}).  In this {\it Letter} we show that although neither
of the latter two mechanisms works on its own, together they have
novel implications for the thermodynamics of the ICM.

The central parts of clusters are unstable to a convective instability
driven by the anisotropic flow of heat along magnetic field lines (the
HBI; \citealt{quat08}).  Simulations of the HBI show that it saturates
by preferentially reorienting the magnetic field lines to be
perpendicular to the temperature gradient, thus reducing the effective
radial conductivity of the plasma \citep[][hereafter
PQS]{pq08,bog09,pqs09}.  This exacerbates the cooling catastrophe by
making it difficult to tap into the thermal bath at large radii,
particularly for clusters with low central entropies and short central
cooling times.

Previous simulations of the HBI in clusters have focused on idealized
problems in which the HBI was the primary source of turbulence
(\citealt{sharma09} also studied the role of convection driven by
cosmic-rays).  There are, however, many additional sources of
turbulence in clusters, including major mergers, the motion of
galaxies through the ICM (galaxy wakes), and AGN jets and bubbles.
Cosmological hydrodynamic simulations of cluster formation find that
turbulence can contribute $\sim 1$--10\% of the total pressure even in
relaxed clusters, with the turbulent pressure declining at small radii
towards the cluster core \citep{lkn09}.
  
In this {\em Letter}, we consider a simple model for the interplay
between turbulence, anisotropic thermal conduction, and radiative
cooling in galaxy cluster cores: we externally ``stir'' the ICM in our
previous global cluster core simulations (e.g., PQS) in order to mimic
the effects of the various sources of turbulence noted above. The
limitations of this approach are discussed in \S \ref{sec:disc}.  

Near the completion of this work, \citet{rus10} presented results
similar to those found here using independent numerical techniques and
cluster models.


\section{Methods}\label{sec:methods}
We solve the equations of MHD using the {\tt Athena} MHD code
\citep{gs08,sg08}, with the addition of anisotropic thermal conduction
\citep{ps05,sh07} and optically thin cooling (see eqs [8]--[12] of
PQS).  In particular, the conductive heat flux is given by
$\boldsymbol{Q} = -
\kappa_{\textrm{Sp}}\boldsymbol{\hat{b}\hat{b}}\cdot\boldsymbol{\nabla}T$,
where $\kappa_{\textrm{Sp}}$ is the Spitzer thermal conductivity and
$\boldsymbol{\hat{b}}$ is a unit vector along the magnetic field.  We
use the \citet{tn01} cooling curve and a temperature floor of
$T=0.05\,\textrm{keV}$, below which UV lines become important.

Our initial condition is a cluster inspired to resemble that of Abell
2199 as observed in \citet{johnstone02}.  We use a static NFW
gravitational potential with a scale radius of $r_s =
390\,\textrm{kpc}$ and a mass of $M_0 = 3.8 \times 10^{14} M_{\odot}$.
The simulations are carried out on a Cartesian grid in a computational
domain that extends from the center of the cluster out to 240 kpc.
The simulations are ($128^3$), corresponding to a resolution of 3.4
kpc.  Resolution studies indicate that our conclusions are not
sensitive to resolution.  We begin with an ICM that is in both
hydrostatic and thermal equilibrium, with conduction balancing
cooling; simulations without initial thermal equilibrium showed large
thermal transients.  The resulting model cluster has a central
temperature and electron density of $\simeq 1.3$ keV and $\simeq 0.04
\;\textrm{cm}^{-3}$, respectively, and a temperature and density of 5
keV and $2\times 10^{-3}\;\textrm{cm}^{-3}$ near 200 kpc.  The
magnetic field is initially tangled, with $\la |B| \ra = 10^{-9}$ G
and a Kolmogorov power spectrum.  Further details of our initial
conditions can be found in sections 3--4 of PQS.

Our model clusters do not include the turbulence that would
realistically be generated by the hierarchical growth of structure or
a central AGN.  To crudely account for this, we continuously add an
additional random velocity to our cluster models.  We drive the
velocity fields in Fourier space using the methods described in
\citet{ls09}. For a given driving length scale $L$, we drive
velocities with a flat spectrum in Fourier space on scales
corresponding to $L\pm 10$ kpc.  We clean the spectrum so that the
motions are incompressible and Fourier transform to real space
normalizing to the desired level of turbulence.  This results in rms
velocities, $\delta v$, that are uniform throughout the cluster.  Note
that our stirring is statistically steady in both time and space.
This is not necessarily a good approximation in clusters---we return
to this point in \S \ref{sec:disc}.

Our fiducial turbulence parameters are $\delta v \sim 50$--$100 \, \kms$
and $L \sim 40$--$100$ kpc.  This corresponds to turbulence that
contributes a few \% of the total pressure in the cluster core.  The strength
of the turbulence produced by galaxies moving through the ICM can be
estimated by calculating the total power dissipated by dynamical
friction (e.g., eqn. [4] of \citealt{kim05}).  Assuming that this energy
is ultimately dissipated via a turbulent cascade, we find
\begin{equation}
  \delta v \sim c_s\left(\frac{GM_g}{c_s^2R}\right)^{2/3}\left(\frac{6N_gL}{R}\right)^{1/3},
\label{eqn:turbv}
\end{equation}
where $M_g$ is the mass of a typical galaxy, $N_g$ is the number of
galaxies, $R$ is the size of the region of interest, and $c_s$ is the
sound speed of the ICM.  For five $10^{11}M_{\odot}$ galaxies within
200 kpc and a turbulent scale of $L \sim 40$ kpc, equation
(\ref{eqn:turbv}) implies $\delta v\sim 0.08 c_s$, consistent with our
fiducial numbers quoted above.  In reality, the generation of
turbulence by galaxies moving through the ICM will be more subtle,
with some of the energy going into sound waves and gravity waves, and
some being confined to galactic wakes correlated with the orbits of
galaxies.

\section{Results}\label{sec:results}

\begin{figure}[tbp!] 
\epsscale{0.5}
\centering
\includegraphics[clip=true, scale=0.45]{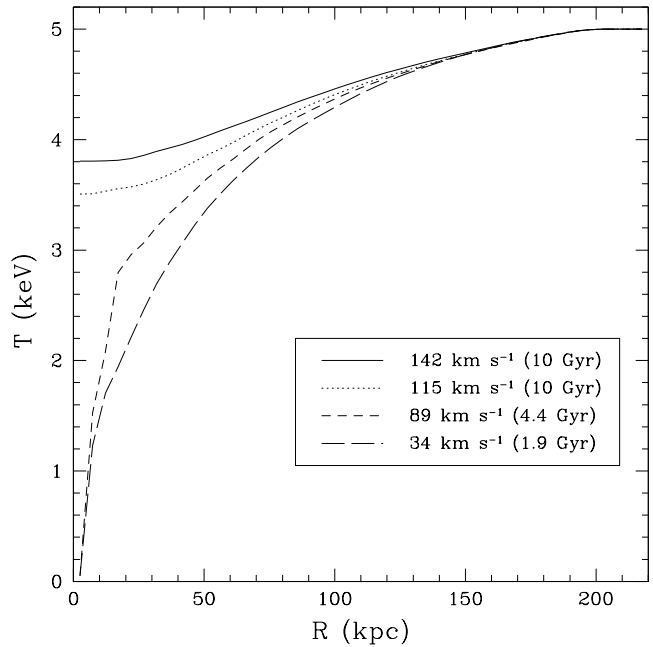}
\caption{Azimuthally-averaged temperature profiles for identical
  cluster cores with different imposed rms turbulent velocities (see
  legend); the driving scale is fixed at $L = 40$ kpc.  A strong
  bimodality in cluster properties results.  Stronger driving ($\delta
  v \gtrsim 100 \, \kms$) leads to a roughly stable thermal profile
  (shown at 10 Gyr).  Somewhat weaker driving leads to a cooling
  catastrophe (shown at the onset of the cooling catastrophe).
  Eqn. \ref{hbi} quantifies the level of turbulent mixing required to
  suppress the HBI and transition from the low temperature, low
  entropy state, to the high temperature, high entropy
  state.} \label{fig:edotvar}
\end{figure}

To illustrate the effect that turbulence can have on the thermal
evolution of galaxy clusters, Figure \ref{fig:edotvar} shows the
late-time azimuthally averaged temperature profile for simulations of
the same cluster with different rms turbulent velocities, $\delta v$,
and thus also different turbulent heating rates $\dot e_{\rm turb}
\simeq \rho \, (\delta v)^3/L$.  For low turbulent velocities, the
evolution is similar to that found previously by PQS: the HBI reorients the magnetic field to be
perpendicular to the radial temperature gradient, shutting off heat
conduction from large radii and thus precipitating a cooling
catastrophe in the cluster core.  The cluster core reaches the
temperature floor at $\simeq 1.9$ Gyr in our lowest $\delta v$
simulation.  For larger $\delta v \gtrsim 100 \, \kms \sim 0.1 \,
c_s$, however, the dynamics changes completely.  The magnetic field
remains relatively isotropic at all times, indicating that the HBI is
no longer acting effectively.  Moreover, the cluster reaches a stable equilibrium, with the temperature profiles shown in
Figure \ref{fig:edotvar} remaining roughly the same for the last
$\simeq 5$--7 Gyr of the simulation.  It is important to stress that
even for $\delta v \sim 100 \, \kms$, the heating rate due to the
turbulence is negligible compared to the cooling rate and thus the
turbulence is energetically unimportant for the thermal properties of
the cluster (\S \ref{sec:disc}).  Note also that the central
temperature increases slightly as the turbulent energy increases in
Figure \ref{fig:edotvar}; we attribute this to the increased advective
(turbulent) heat transport associated with the higher turbulent
velocities.

\begin{deluxetable*}{ccccccccc}[tbp!]
\tablecolumns{9}
\tablecaption{Properties of the Fiducial Simulations\tablenotemark{a} \label{tab:fid}}
\tablewidth{0pt}
\tablehead{
\colhead{Run} &
\colhead{$\la \edot \ra$ ($\power$)} &
\colhead{$L$ (kpc)} &
\colhead{$\delta v$ ($\kms$)} &
\colhead{$\tcool$ (Myr)} &
\colhead{$t_{\rm HBI}$ (Myr)} &
\colhead{$t_{\textrm{eddy}}(L)$ (Myr)} &
\colhead{$t_{\textrm{eddy}}(\lambda_F)$ (Myr)} &
\colhead{$K_0^{\textrm{final}}\, (\kevcm)$}
}
\startdata
A & $7.5\times 10^{-30}$ & 40 & 115 & 400 & 100 & 400 & 360 & 53\\
B & $3.0\times 10^{-30}$ & 100 & 112 & 400 & 100 & 1000 & 490 & 0.06 \\
C & $3.0\times 10^{-29}$ & 100 & 250 & 400 & 100 & 390 & 195 & 110
\enddata
\tablenotetext{a}{Timescales are estimated at the initial time, while
  the turbulent velocity and central entropy are measured in the
  saturated state for runs A \& C, and just before the cooling
  catastrophe for run B. The Field length is estimated near the
  cluster center, while $\edot$ is volume averaged.}
\end{deluxetable*} 

\begin{figure}[] 
\epsscale{0.6}
\centering
\includegraphics[clip=true, scale=0.46]{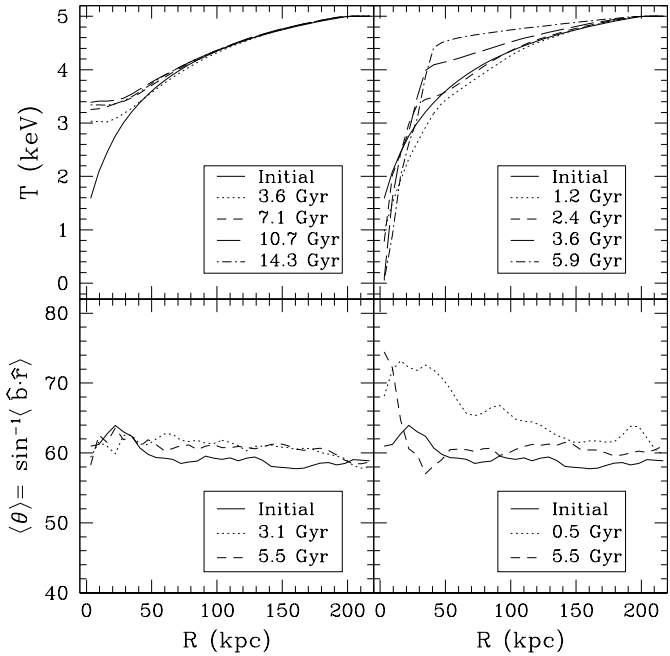}
\caption{Azimuthally-averaged profiles for simulations with imposed
  turbulence having the same $\delta v \simeq 112 \, \kms$, but
  different correlation lengths.  Run A ({\em left}) has $L = 40$ kpc,
  while run B ({\em right}) has $L = 100$ kpc (see Table
  \ref{tab:fid}) \textit{Top:} Temperature profiles. \textit{Bottom:}
  Magnetic field angle relative to the radial: $60\degree$ corresponds
  an isotropically tangled magnetic field and an effective radial
  conductivity $\sim 1/3$ Spitzer.  Run A (\textit{left}), with the
  shorter turbulent mixing time, reaches a stable state averting
  the cooling catastrophe.  In Run B (\textit{right}), the HBI
  persists, shutting off conduction and initiating a cooling
  catastrophe.}  \label{fig:fiducial}
\end{figure}

To quantify in more detail the effect of turbulence on the evolution
of cluster plasmas, Figure \ref{fig:fiducial} shows the temperature
and magnetic field direction as a function of radius at several
different times for two simulations (labeled A and B) whose properties
are summarized in Table \ref{tab:fid} (case C in Table \ref{tab:fid}
is discussed below).  The simulations are again of identical clusters
and both include turbulence with $\delta v \sim 100 \, \kms$, near the
threshold for the transition from stability to instability in Figure
\ref{fig:edotvar}.  The simulations differ in that case A has a
turbulent correlation length of $L = 40$ kpc, while B has $L = 100$
kpc; as a result, simulation A has a heating rate $\dot e_{\rm turb}$
that is 2.5 times higher and an eddy turnover time on scale $L$ that
is $2.5$ times shorter.

Despite their similarities, the top panels of Figure
\ref{fig:fiducial} show dramatic differences in the evolution of the
clusters' radial temperature profile.
Simulation B proceeds with little regard for the turbulent driving.
The HBI reorients the magnetic field, reducing the effective radial
thermal conductivity, and hastening the onset of the cooling
catastrophe at $\simeq 2.2$ Gyr.  On the other hand, simulation A is
dramatically affected by the turbulent driving, just like the high
$\delta v$ simulations in Figure \ref{fig:edotvar}.  In this case the
turbulence effectively shuts off the HBI, with the mean angle of the
magnetic field from radial fluctuating about the isotropic value of
$\la\theta_B\ra\sim 60\degree$.

In addition to suppressing the effects of the HBI, the presence of
``sufficient'' turbulence also appears to modify the thermal stability
of the cluster.  Case A in Figure \ref{fig:fiducial} shows that the
cluster reaches a statistically stable thermal equilibrium that
survives for longer than the age of the universe.  The stability of
the new equilibrium is illustrated by the fact that the cluster's
central temperature undergoes slight oscillations about a new
equilibrium value.

\section{Interpretation and Discussion}\label{sec:disc}

It is important to stress that for all of the simulations presented
here, the heating produced by the turbulence is energetically
subdominant (this is also consistent with the modest dynamical
friction heating in clusters inferred using observed galaxies; e.g.,
\citealt{kim05}).  For simulation A in Figure \ref{fig:fiducial},
e.g., the turbulent heating rate per unit volume at the center of the
computational domain is $\lesssim 2 \%$ of the central cooling rate
throughout the evolution of the cluster.  As a result, the physics
important for the results in Figures \ref{fig:edotvar} \&
\ref{fig:fiducial} includes anisotropic thermal conduction, the HBI,
the thermal instability of cluster plasmas \citep{field65}, and the
mixing produced by the turbulence---but not the heating by such
turbulence.

In general the global thermal instability of cluster plasmas in the
presence of thermal conduction manifests itself as either catastrophic
cooling in the core of the cluster, or overheating and the approach to
an isothermal temperature profile (e.g., \citealt{conroy08}; PQS).
Which of these is realized in a given problem depends in part on the
initial state of the system and boundary conditions.  In cluster
simulations without externally imposed turbulence, the HBI biases the
nonlinear evolution of the thermal instability towards the cooling
catastrophe by thermally decoupling the core from larger radii.

Our simulations show that if turbulence is sufficiently strong in
cluster cores it can effectively shut off the HBI, leaving the
magnetic field tangled and relatively isotropic, and the cluster with
a quasi-steady, not-quite-isothermal temperature profile
(Figs. \ref{fig:edotvar} \& \ref{fig:fiducial}).  Quantitatively,
turbulence with $\delta v \gtrsim 100 \, \kms$ or a Mach number
$\gtrsim 0.1$ appears sufficient.  More precisely, we believe that
the critical criterion is \citep{sharma09b} \be t_{\rm eddy}(L) \simeq
{L \over \delta v} \lesssim \xi \, t_{\rm HBI} \simeq \xi \, \left(g {d\ln
    T \over dr} \right)^{-1/2} \label{hbi} \ee where $t_{\rm eddy}(L)$
is the timescale for the turbulence to mix the plasma at the outer
scale, $t_{\rm HBI}$ is the HBI growth time, $g$ is the local
gravitational acceleration in the cluster, and $\xi$ is a dimensionless
constant that must be determined from the simulations.  Note that
because the mixing timescale is smaller on smaller scales in a
Kolmogorov cascade, the timescale inequality in equation (\ref{hbi})
is the most difficult to satisfy at the outer scale.

Figure \ref{fig:fiducial} demonstrates explicitly that a given value
of $\delta v$ is {\em not} in fact sufficient to suppress the HBI and
halt the cooling catastrophe.  Rather, this only occurs in run A,
which has a smaller correlation length $L$ and shorter mixing time
than run B (Table \ref{tab:fid}). Table \ref{tab:fid} shows the
properties of a third simulation not in Figure \ref{fig:fiducial}, in
which the correlation length is the same as in run B ($L = 100$ kpc),
but the turbulent velocity is larger.  This combination again
satisfies equation (\ref{hbi}) and so the evolution is qualitatively
similar to run A.  Because the HBI growth time is $\sim 100$ Myr at
$\sim 50$ kpc in these models, our numerical results correspond to
$\xi\simeq 5$--8 in eq. (\ref{hbi}).

In addition to its effects on the HBI, turbulent mixing can also
significantly modify the thermal stability of cluster plasmas.
Independent of turbulent mixing, thermal instability is stabilized on
small scales (along the magnetic field) by thermal conduction.  The
critical length-scale below which conduction suppresses thermal
instability (the Field length) is given by \begin{equation} \lambda_F
  \equiv \left[\frac{4 \pi^2
      T\kappa(T)}{n_en_p\Lambda(T)}\right]^{1/2} \approx
  53\,\textrm{kpc}\,\left[\frac{K}{15\;\textrm{keV}\,\textrm{cm}^2}\right]^{3/2},
\label{eqn:fieldlength}
\end{equation}
where $\Lambda(T)$ is the cooling function and we have used the full
Spitzer conductivity in the final expression; in the second equality
we have also assumed for simplicity that the cooling is pure
bremsstrahlung so that $\lambda_F$ can be expressed solely in terms
of the entropy $K = Tn_e^{-2/3}$ \citep{donahue05}.

Absent turbulence, fluctuations with length-scales $\gtrsim \lambda_F$
are unstable on a cooling time $\tcool$.  If, however, turbulent
mixing is sufficiently rapid, i.e., if \be t_{\rm eddy}(\lambda_F)
\lesssim \tcool,
\label{TI}
\ee then turbulent mixing will suppress the thermal instability on the
scale of the Field length, although larger length-scale fluctuations
may remain unstable; local simulations of thermal instability in the
presence of background turbulence confirm this intuition (these will
be presented elsewhere).  Table \ref{tab:fid} shows that for both runs
A and B in Figure \ref{fig:fiducial}, $t_{\rm eddy}(\lambda_F) \sim
\tcool$; that is, the modest turbulence levels considered here are
capable of significantly changing the dynamics of the thermal
instability in cluster plasmas.  One extreme limit of this is the
possibility that turbulent mixing of hot gas from large radii with
cooler gas from small radii can help prevent the cooling catastrophe
\citep{markevitch10}.  In our stable simulations, however, (e.g., Case
A) this is not realized: thermal conduction (not turbulent mixing)
provides the dominant source of heating at small radii.  The key role
of the turbulent mixing is that it suppresses the HBI, isotropizes the
magnetic field, and helps suppress the thermal instability by mixing
the plasma before it can cool.  This dynamics is qualitatively
analogous to the critical role that turbulence plays in mixing and
disrupting laminar conductive flames in the combustion and Type Ia
supernova contexts \citep[e.g.,][respectively]{ peters00, woosley07}.

In our simulations, the interaction between turbulence, the HBI, and
cooling leads to a strong bimodality in the cluster properties (e.g.,
temperature profiles).  Figure \ref{fig:edotvar} shows that runs with
moderately strong turbulence (satisfying eqn. [\ref{hbi}]) reach a
quasi-stable thermal equilibrium averting the cooling catastrophe.  By
contrast, runs with slightly weaker turbulence---$\delta v$ smaller
by just $\sim 25 \%$---progress to a cooling catastrophe on a
timescale as short as a few central cooling times (much like the pure
HBI simulations of PQS and \citealt{bog09}).

It is tempting to relate this behavior to the observed variety in
galaxy cluster properties.  Observationally, clusters show a
bimodality in their central gas entropies and cooling times, with
lower entropy clusters preferentially having more star formation and
H$\alpha$ emission, and more powerful AGN \citep{voit08, cav09}; the
transition occurs at $\simeq 30 \;\kevcm$.  This bimodality is closely
related to the well-known fact that clusters come in both cool core
and non cool core varieties (see \citealt{hudson09}).

The observed bimodality in cluster properties is not well
understood. \citet{burns08} argued that early major mergers could
prevent the formation of cool core clusters.  Alternatively, \citet{gor08} showed using both 1D time dependent
models and a global stability analysis that the combination of AGN
feedback and scalar conduction can produce a bimodal population of
stable cluster models; in their models AGN heating largely balances
cooling in lower entropy clusters while conduction is more important
in higher entropy clusters.  They further suggested that AGN feedback
could transition clusters from low to high entropy \citep{guo09}.

Our simulations demonstrate that a low entropy cool-core cluster can
transition to a significantly higher entropy state: run A initially
has a central entropy of $K_0\approx 11\;\kevcm$ while its final
central entropy is $K_0\approx 57\;\kevcm$.  Run C is the same as run
B but with a higher $\delta v$; its final central entropy is $K_0\sim
110 \; \kevcm$.  This increase in central entropy is a consequence of
runaway conductive heating at the roughly fixed pressure required for
hydrostatic equilibrium.  Physically, such a transition could occur if
a cluster initially had little turbulence and inefficient conduction
(because of the HBI), but was stabilized by a central AGN.  The sudden
onset of turbulence satisfying equation (\ref{hbi})---produced by an infalling galaxy or the AGN---would isotropize the magnetic
field and suppress the HBI.  The cluster would then evolve as in
Figure \ref{fig:fiducial} ({\em left panel}; case A) to a higher
entropy state.  If the turbulence in the cluster core later died away,
the HBI would rapidly rearrange the magnetic field
(Fig. \ref{fig:edotvar}), leading to cooling of the core after $\sim
1$ Gyr and (by assumption) increased AGN activity that would again
stabilize the cluster in a cool-core state.  This demonstrates that
modest levels of turbulence in cluster cores ($\delta v \sim 100 \;
\kms$; eqn. [\ref{hbi}]) can have a dramatic affect on their thermal
evolution.  Note also that the energy required to generate turbulence
capable of suppressing the HBI and thermal instability is far less
than that required to directly increase the entropy of the cluster (as
in \citealt{guo09}).

Our calculations are based on an overly-simplified treatment of
turbulence in galaxy cluster plasmas.  Real turbulence in clusters is
likely to be more intermittent in space and time than our model (\S
\ref{sec:methods}), depending on, e.g., the distance to an AGN jet or
a galactic wake.  Waves generated by AGN jets and bubbles and/or
galaxies moving through the ICM may produce reasonably volume-filling
turbulence in cluster cores, but this needs to be studied in
detail. Provided that the turbulence is replenished on $t_{\rm
  cool} \sim 0.1$--1 Gyr, our results will be relatively unchanged.
Otherwise, the HBI and thermal instability will proceed unchecked.
This temporal and spatial intermittency of turbulence in cluster
cores may ultimately prove to be a positive feature, not a ``bug,'' of
our model: as described above, modest changes in the level of
turbulence in clusters can produce rapid and dramatic changes in the
thermal structure and stability of the cluster, to the point of
initiating transitions from low to high entropy states (and
vice-versa).  Overall, the subtle interaction between turbulence, the
HBI, and cooling in galaxy cluster cores has a surprisingly large
impact on the thermal properties of the ICM.  The critical role of the
turbulence is not the small amount of turbulent energy dissipated
(which is $\ll$ the cooling luminosity); rather, it is the fact that
turbulent mixing can suppress both the HBI and the thermal instability
in cluster cores (see eqs. [\ref{hbi}] and [\ref{TI}], respectively).

\acknowledgements Support for I.~J.~.P and P.~S. was provided by NASA
through Chandra Postdoctoral Fellowship grants PF7-80049 and PF8-90054
awarded by the \ch X-Ray Center, which is operated by the Smithsonian
Astrophysical Observatory for NASA under contract NAS8-03060.
E.~Q. was supported in part by the David and Lucile Packard
Foundation, NSF-DOE Grant PHY-0812811, and NSF ATM-0752503.  Computing
time was provided by the National Science Foundation through the
Teragrid resources located at the National Center for Atmospheric
Research and the Pittsburgh Supercomputing Center.
\bibliography{ms}
\end{document}